\documentclass{article}
\usepackage{spconf,amsmath,graphicx}
\usepackage{bm}
\usepackage{threeparttable}
\usepackage{multirow}
\usepackage{booktabs}
\usepackage{color}
\title{Invertible Mosaic Image Hiding Network for Very Large Capacity Image Steganography}
\name{Zihan Chen$^1$, Tianrui Liu$^1$, Jun-Jie Huang$^1$, Wentao Zhao$^{1\dagger}$, Xing Bi$^2$ and Meng Wang$^3$}
\address{
$^1$College of Computer Science and Technology, National University of Defense Technology, China\\
$^2$College of System Engineering, National University of Defense Technology, China\\
$^3$School of Computer Science and Information Engineering, Hefei University of Technology, China
\thanks{$^\dagger$Corresponding Author}
}
\begin{document}
%\ninept
%s
\maketitle
\begin{abstract} 
The existing image steganography methods either sequentially conceal secret images or conceal a concatenation of multiple images. In such ways, the interference of information among multiple images will become increasingly severe when the number of secret images becomes larger, thus restrict the development of very large capacity image steganography. 
In this paper, we propose an Invertible Mosaic Image Hiding Network (InvMIHNet) which realizes very large capacity image steganography with high quality by concealing a single mosaic secret image. InvMIHNet consists of an Invertible Image Rescaling (IIR) module and an Invertible Image Hiding (IIH) module. 
The IIR module works for downscaling the single mosaic secret image form by spatially splicing the multiple secret images, and the IIH module then conceal this mosaic image under the cover image. 
The proposed InvMIHNet successfully conceal and reveal up to 16 secret images with a small number of parameters and memory consumption.
Extensive experiments on ImageNet-1K, COCO and DIV2K show InvMIHNet outperforms state-of-the-art methods in terms of both the imperceptibility of stego image and recover accuracy of secret image.
\end{abstract}
\begin{keywords}
Image steganography, Image rescaling, Invertible neural networks
\end{keywords}
\section{Introduction}
\label{sec:intro}

Image steganography~\cite{baluja2017hiding,Zhu2018HiDDeNHD,weng2019high,Kadhim2019ComprehensiveSO} refers to hiding secret images in a cover image and the resultant stego image needs to be visually similar to the cover image. 
Meanwhile, the generated stego images are required to contain sufficient information for reconstructing the secret images with accurate details.
Traditional image steganography methods hide information behind the cover image in image domain or transform domain using handcrafted features~\cite{Chan2004HidingDI,nguyen2016adaptive,atawneh2017secure}, and usually suffer from low hiding capacity of $0.2 \sim 4$ bits per pixel (bpp). Recent learning-based image steganography methods improve the hiding capability with deep neural networks by learning from external training data~\cite{baluja2017hiding,Zhu2018HiDDeNHD,zhang2020udh}.
Since they treat concealing and revealing as two individual tasks, the hiding capacity and reconstructed performance are limited. 
To have tighter connections between concealing and revealing processes, more recent works introduce Invertible Neural Networks (INNs)~\cite{dinh2014nice, dinh2016density, kingma2018glow, xiao2020invertible, huang2021winnet} into image steganography and learn a reversible mapping from input manifold to the output with shared parameters leading to improved hiding capacity. 
Lu \textit{et al.}~\cite{Lu2021LargecapacityIS} proposed Invertible Steganography Network (ISN) for large capacity image steganography by utilizing invertibility of the Invertible Neural Networks to construct intrinsically coupled hiding and revealing networks. They demonstrated that ISN is able to hide up to 5 secret images. 
Jing \textit{et al.}~\cite{hinet} proposed HiNet to incorporate INNs with wavelet transform for image hiding, leading to improved concealing quality and recovery accuracy. 
Guan \textit{et al.}~\cite{guan2022deepmih} further improved HiNet and proposed Deep Invertible Network for Multiple Image Hiding (DeepMIH) by cascading multiple invertible hiding networks.

The existing learning-based methods can now successfully conceal and reveal up to five color secret images, either by simply concatenating multiple secret images as a whole input or by cascading multiple concealing networks.
However, these methods would encounter perceptible color distortions on the recovered secret images  when trying to conceal more secret images. Moreover, the memory occupation and computational cost will increase with the number of images to be hidden.
We argue that the current tactics fail to fully exploit the spatial redundancy of the secret images and are suboptimal. This fundamentally limits further improvement on hiding capacity and recovery quality.

% ---------------------------------------------------------------------------------
\begin{figure*}[t]
    \centering
    \includegraphics[width=0.95\linewidth]{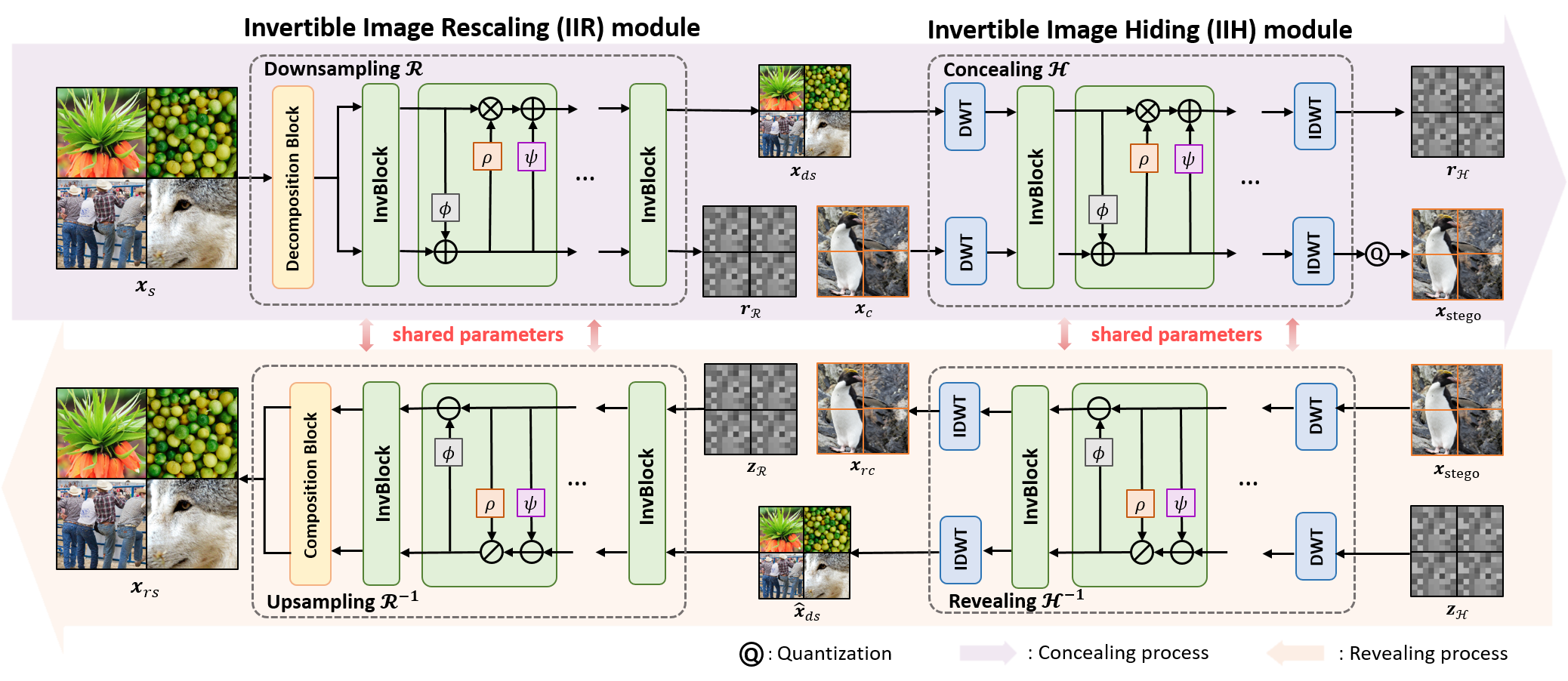}
    \caption{Overview of the proposed Invertible Mosaic Image Hiding Network (InvMIHNet) with $N=4$ secret images. The secret images $\bm{x}_{s}$ are firstly downscaled by a factor of $2\times 2$ using the forward pass of the Invertible Image Rescaling (IIR) module and form a mosaic secret image. 
    The cover image $\bm{x}_{c}$ and the MSI $\bm{x}_{ds}$ are then input to the forward pass of the Invertible Image Hiding (IIH) module which outputs the stego images $\bm{x}_{\text{stego}}$ and an image-agnostic components $\bm{r}_{\mathcal{H}}$. The recovered secret images $\bm{x}_{rs}$ are then obtained by sequentially applying the reverse pass of IIH and IIR modules with revealing and upscaling. }
    \label{fig:overview}
\end{figure*}
% ---------------------------------------------------------------------------------

In this paper, we propose a novel Invertible Mosaic Image Hiding Network (InvMIHNet) for very large capacity image steganography and show that we can, for the first time, improve the hiding capacity up to 16 color secret images.
The proposed InvMIHNet consists of an Invertible Image Rescaling (IIR) module and an Invertible Image Hiding (IIH) module.
During the image concealing process, the secret images are first rescaled to a compact representation which is of the same spatial size as the cover image by the IIH module. We then form a single mosaic secret image by spatially splicing the downscaled secret images. The mosaic secret image will be concealed in the cover image by the IIR module. 
Owing to the invertibility of Invertible Neural Networks, during the revealing process, the reverse pass of the IIR module can faithfully recover the rescaled secret images which are then upscaled to the original resolution by the reverse pass of the IIH module. 
In this way, the spatial redundancy of the secret images can be explicitly exploited, and the information interference from different secret images and color distortions are minimized, leading to a significant improvement on hiding capacity and recovery quality.

\section{Proposed Method}
\label{sec:proposed_InvMIHNet}

The overview of the proposed Invertible Mosaic Image Hiding Network (InvMIHNet) is shown in Fig.~\ref{fig:overview}. Here, we take the case of $N=4$ secret images for illustration\footnote{For illustration, we assume that all the secret images and cover image are of the same spatial size here.}. 
The $N=m \times n$ secret images are first downscaled by a factor of $m\times n$ using the Invertible Image Rescaling (IIR) module, obtaining downscaled secret images together with image-agnostic components $\bm{r}_{\mathcal{H}}$. The downscaled secret images are then spatially spliced to form a mosaic secret image (MSI) $\bm{x}_{ds}$.
The information of $\bm{x}_{s}$ is faithfully retained in $\bm{x}_{ds}$ and is transformed to a form which can be easily concealed in the cover image $\bm{x}_{c}$.
The Invertible Image Hiding (IIH) module then takes $\bm{x}_{ds}$ and $\bm{x}_{c}$ as inputs and 
% And the other input is the cover image $\bm{x}_{c}$ and 
results in the stego image $\bm{x}_{\text{stego}}$ and an image-agnostic latent variable $\bm{r}_{\mathcal{H}}$.
The revealing process is then naturally the inverse operations of the IIH and IIR modules which are enabled by the invertibility of INNs.

\noindent\textbf{Invertible Blocks (InvBlocks)} are utilized to construct Invertible Neural Networks (INNs)~\cite{dinh2014nice, dinh2016density, kingma2018glow, xiao2020invertible} whose forward pass and reserve pass share the same sets of parameters and are bijective functions. 
% Therefore, INNs can be used to learn non-linear transformations, moreover its inverse function can be explicitly constructed.
Let us denote with $\bm{x}_{l}^{i}$ and $\bm{x}_{h}^{i}$ the input of the $i$-th InvBlock, and $\bm{x}_{l}^{0},\bm{x}_{h}^{0}=\mathcal{D}({\bm{x}_{ds}})$ where $\mathcal{D}(\cdot)$ is the decomposition block depending on the specific task. 
The forward process of the $i$-th InvBlock can be expressed as:
\begin{equation}
    \begin{cases}
    \bm{x}_{l}^{i} &= \bm{x}_{l}^{i-1}+{\phi}({\bm{x}_{h}^{i-1}}),\\
    \bm{x}_{h}^{i} &= \bm{x}_{h}^{i-1} \odot \exp \left(\rho \left (\bm{x}_{l}^{i}\right) \right) + {\psi}(\bm{x}_{l}^{i}),
    \end{cases}
\end{equation}
where $\bm{x}_{l}$ and $\bm{x}_{h}$ denote the low- and high-frequency contents, $\odot$ denotes element-wise multiplication, $\phi(\cdot), \rho(\cdot)$ and $\psi(\cdot)$ represent dense networks \cite{huang2017densely}. 
Hence, the reverse process of the $i$-th InvBlock can be expressed as:
\begin{equation}
    \begin{cases}
        \bm{x}_{h}^{i-1} &= \left (\bm{x}_{h}^{i}-{\psi}\left (\bm{x}_{l}^{i} \right) \right) \odot \exp \left(-\rho \left (\bm{x}_{l}^{i} \right) \right),\\
        \bm{x}_{l}^{i-1} &= \bm{x}_{l}^{i}-{\phi}({\bm{x}_{h}^{i-1}}).
    \end{cases}
\end{equation}

\noindent\textbf{Invertible Image Rescaling (IIR) module}: IIR module performs reversible mosaic secret image rescaling. 
When concealing $N=m \times n$ secret images in the cover image, the secret images $\bm{{x}}_{s}$ are spatially downscaled by a factor of $m \times n$ and is transformed to form a mosaic secret image (MSI) $\bm{x}_{ds}$ so that essential information of the secret images can be easily concealed in the cover image.
The forward pass of the IIR module performs downscaling in the concealing process and consists of a decomposition block $\mathcal{D}(\cdot)$ and $R$ forward InvBlocks. One the opposite, the reverse pass of the IIR performs upscaling in the revealing process and consists of $R$ reverse InvBlocks and a composition block $\mathcal{D}^{-1}(\cdot)$.
The decomposition block $\mathcal{D}(\cdot)$ transforms the secret image $\bm{x}_{s} \in \mathcal{R}^{C \times Hm \times Wn}$ into $\mathcal{D}(\bm{x}_{s}) \in \mathcal{R}^{mnC \times H \times W}$ 
% with a convolution layer whose kernel is parameterized by Cayley transform 
and then splits it into high-frequency sub-bands $\bm{x}_{h} \in \mathcal{R}^{(mn-1)C \times H \times W}$ and low-frequency sub-band $\bm{x}_{l} \in \mathcal{R}^{C \times H \times W}$. The composition block $\mathcal{D}^{-1}(\cdot)$ reversely merges the low-frequency sub-band and high-frequency sub-bands, and then the merged contents are transformed back to the original shape. 

\noindent\textbf{Invertible Image Hiding (IIH) module}: In the forward concealing process, IIH module aims to concealing the MSI $\bm{x}_{ds}$ into the cover image $\bm{x}_{c}$; and in the reverse recealing pass, it works for revealing MSI from the stego image.
The IIH module consists of a DWT/IDWT block and $G$ InvBlocks. DWT Block splits the input images into low-frequency and high-frequency sub-bands and can facilitate the secret information concealing in the cover image patch.
By default, Haar Wavelet transform is applied. 
In the forward pass, MSI $\bm{x}_{ds}$ and cover image $\bm{x}_{c}$ are first decomposed by DWT block and then input into the $G$ InvBlocks and IDWT block to generate a stego image and an image-agnostic variable. The stego image will then pass a quantization block to clamp the pixels values to integers within the range of $[0, 255]$.
In the reserve pass, the stego image and a sampled random variable $z_{\mathcal{H}}$ are input to IIH module to restore the cover image and MSI which will then be upscaled by the reverse pass of IIR module to obtain the final recovered secret image $\bm{x}_{rs}$.

\noindent\textbf{Training Details:}
The learning objective involves minimizing the restoration error between the secret image and the recovered secret image, and minimizing the concealing difference between the cover image and the stego image. Therefore, the loss function can be expressed as:
\begin{equation}
    \begin{aligned}
        \mathcal{L} = & \mathcal{JS}(p(\bm{x}_{s}), p(\bm{x}_{rs}))+ \lambda_{1} \left\|\bm{x}_{s}-\bm{x}_{rs} \right\| \\
        &+ \lambda_{2} \left \| \bm{x}_{ds}-\bm{x}_{\text{ref}} \right \|^{2}_{2} + \lambda_{3} \left \| \bm{x}_{ds}-\bm{\hat{x}}_{ds} \right \|^{2}_{2}\\
        &+ \| \bm{x}_{c} - \bm{x}_{\text{stego}}  \|^{2}_{2} 
        +  \| \bm{x}_{c(ll)}-\bm{x}_{\text{stego}(ll)} \| ^{2}_{2},
    \end{aligned}
\end{equation}
where $\mathcal{JS}$ denotes Jensen–Shannon divergence which is set to match the distributions of secret images and recovered secret images, 
$p(\cdot)$ represents the data distribution, $\bm{x}_{\text{ref}}$ denotes Bicubic downscaled secret images, $\{\lambda_i\}_{i=1}^3$ denote the regularization parameters and the subscript $ll$ denotes the low-frequency component of the wavelet transform.
% due to the secret information is mainly concealed in the more imperceptible high-frequency sub-bands.

\section{Experiments}
\label{sec:experiments}
\subsection{Dataset and Settings}
The DIV2K~\cite{agustsson2017ntire} dataset is used for training, which  contains 800 training images.
The testing dataset includes 100 test images of resolution 1024 $\times$ 1024 from the DIV2K validation  dataset ~\cite{agustsson2017ntire}, 1,000 images of resolution 256$\times$256 randomly select from ImageNet~\cite{russakovsky2015imagenet} and 5,000 images of resolution 256$\times$256 from the test dataset of COCO~\cite{lin2014microsoft}.

The experiments were performed on a computer with a single NVIDIA RTX 3090 GPU with 24 GB memory. The stochastic gradient descent with Adam optimizer~\cite{kingma2014adam} is used for training with the initial learning rate $2\times10^{-4}$ which is halved every $10K$ iterations. The training patch size is of 144$\times$144.
The regularization parameters are empirically set to $\lambda_{1}=1$, $\lambda_{2}=4$ and $\lambda_{3}=5$. 
By default, the number of IIR module and IIH module is set to 8 and 16, respectively.

We train InvMIHNet in a two-stage manner.
In the first stage, we perform warm up training on IIR module and IIH module separately for $30K$ iterations. 
In the second stage, we jointly train InvMIHNet for another $20K$ iterations.

% ------------------------------------------------------------------------------------------
\begin{table}[t]
  \centering

  \caption{The experimental results of hiding 4 images of ISN, DeepMIH and InvMIHNet on DIV2K, COCO and ImageNet.}
  
  \small
  % \begin{threeparttable}
  \scalebox{0.9}{
    \begin{tabular}{l|cc|cc|cc}
    \hline
    \multirow{3}[0]{*}{Methods} & \multicolumn{6}{c}{Cover / Stego} \\
\cline{2-7}          & \multicolumn{2}{c|}{DIV2K}    & \multicolumn{2}{c|}{COCO}     & \multicolumn{2}{c}{ImageNet} \\
\cline{2-7}          & PSNR   & SSIM   & PSNR   & SSIM   & PSNR   & SSIM  \\
    \hline
    \hline
    ISN~\cite{Lu2021LargecapacityIS}   & 32.90  & 0.896 & 34.96 & \textbf{0.946} & 32.48 & 0.908 \\
    DeepMIH~\cite{guan2022deepmih}  & 33.94 & 0.904  & 30.93  & 0.873  & 34.29 & 0.924 \\
    InvMIHNet & \textbf{37.66} & \textbf{0.945} & \textbf{34.98} & 0.932 &  \textbf{36.86} & \textbf{0.948}  \\
    \hline
    \hline
    \multirow{3}[0]{*}{Methods} & \multicolumn{6}{c}{Secret / Recovery} \\
\cline{2-7}          & \multicolumn{2}{c|}{DIV2K}    & \multicolumn{2}{c|}{COCO}     & \multicolumn{2}{c}{ImageNet} \\
\cline{2-7}          & PSNR   & SSIM   & PSNR   & SSIM   & PSNR   & SSIM \\

    \hline
    \hline
    ISN~\cite{Lu2021LargecapacityIS}   & 27.65 & 0.815 & 26.66  & 0.823  & 27.93  & 0.826 \\
    DeepMIH~\cite{guan2022deepmih}  & 33.13  & 0.930  & 29.04  & 0.861  & 32.40  & 0.913 \\
    InvMIHNet & \textbf{33.36} & \textbf{0.939} & \textbf{29.11} & \textbf{0.875} & \textbf{32.90} & \textbf{0.931}\\
    \hline
    \end{tabular}%
    }
  \label{tab:conceal_four_2}%
\end{table}%
% ------------------------------------------------------------------------------------------
% ---------------------------------------------------------------------------------
\begin{figure}[t]
    \centering
    \includegraphics[width=0.88\linewidth]{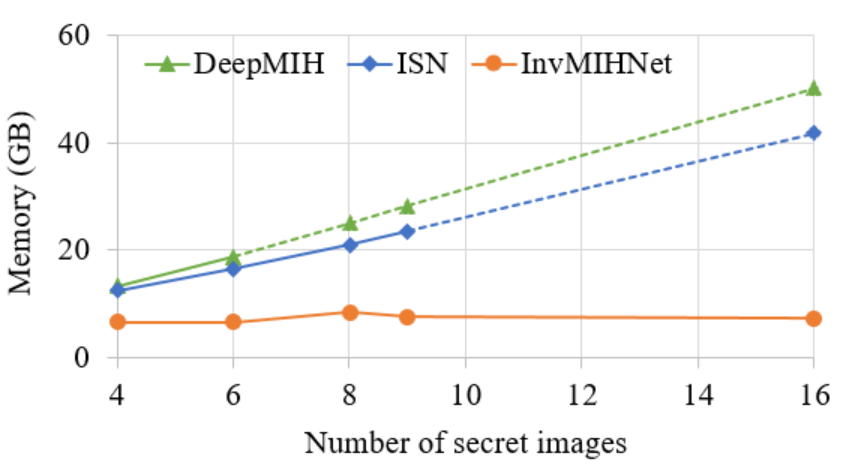}
    \caption{Comparison of memory consumption of secret images $N \in [4,16]$. The dashed lines represent the linear interpolated values, since these models are out of the memory\protect\footnotemark.}
    \label{fig:memory_consumption}
\end{figure}
% ---------------------------------------------------------------------------------

\footnotetext{NVIDIA RTX 3090 GPU with 24 GB memory}

\subsection{Evaluation Results}
DeepMIH~\cite{guan2022deepmih} and ISN~\cite{Lu2021LargecapacityIS} are state-of-the-art large-capacity steganography methods based on Invertible Neural Networks and are included for comparison.
For fair comparison, we train and test all methods on the same datasets.

% ---------------------------------------------------------------------------------
\begin{figure}[t]
    \centering
    \includegraphics[width=0.9\linewidth]{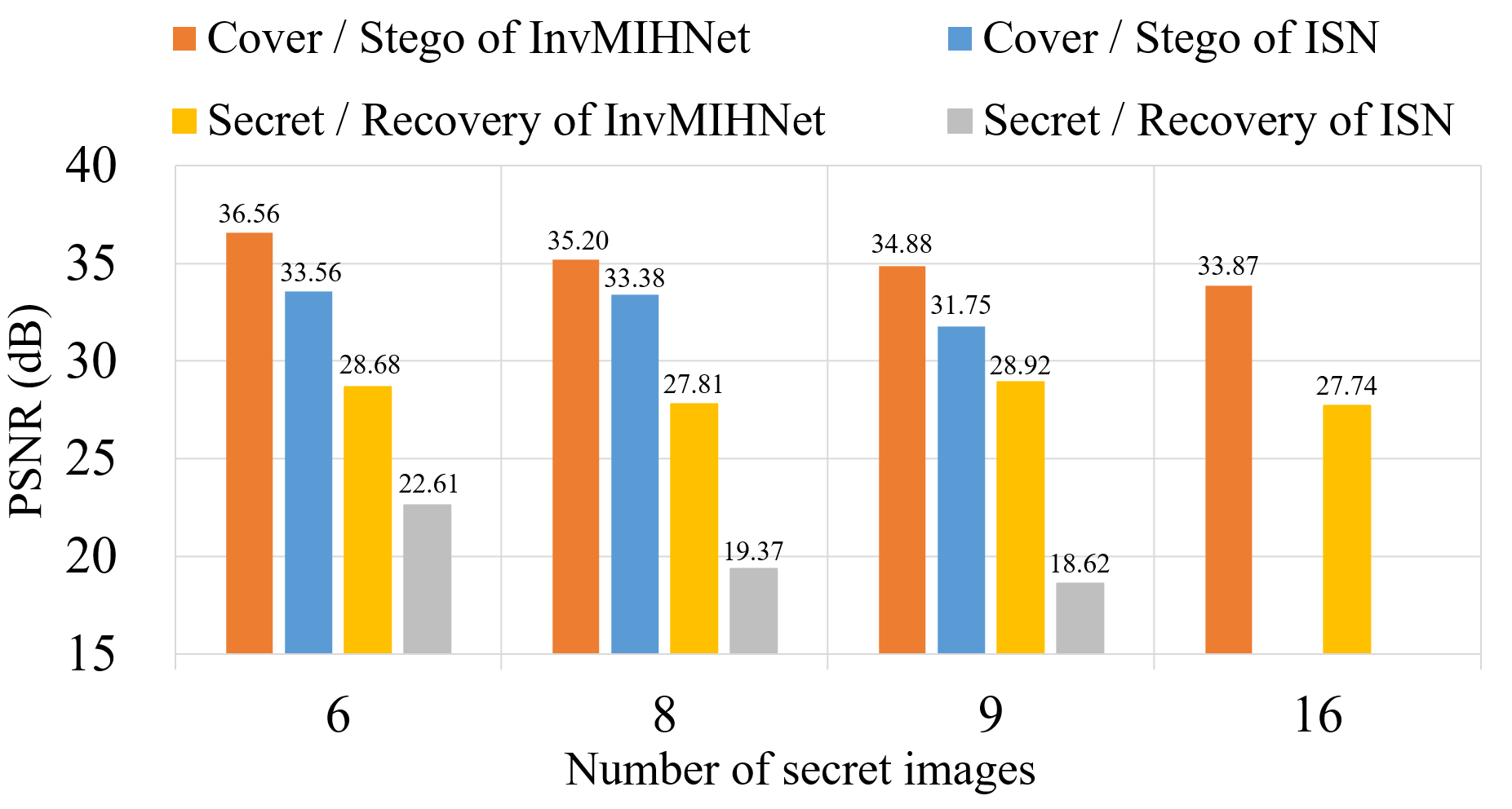}
    \caption{Experimental results on hiding $N (N>4)$ secret images of ISN and InvMIHNet.}
    \label{fig:mult_image}
\end{figure}
% ---------------------------------------------------------------------------------

% ---------------------------------------------------------------------------------
\begin{figure}[t]
    \centering
    \includegraphics[width=1.0\linewidth]{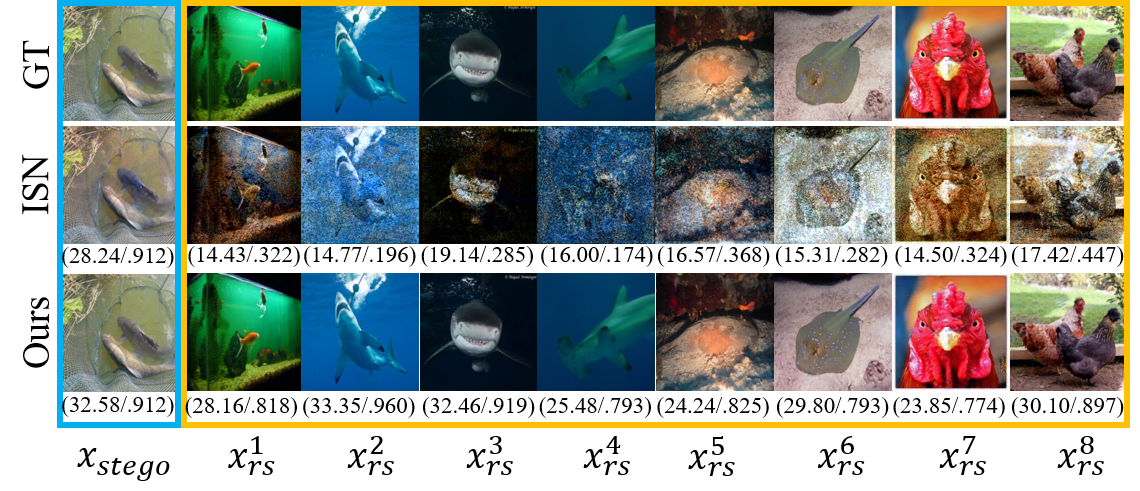}
    \caption{Visualizations on hiding 8 images of ISN and InvMIHNet on ImageNet.}
    \label{fig:8images}
\end{figure}%updated
% ---------------------------------------------------------------------------------

%------------------------------------------------------------------------------
\begin{table}[t]
  \centering
  \small
    \caption{Ablation study on the number of InvBlocks of IIR and IIH modules on hiding 4 secret images.}
    \small
    \begin{tabular}{cc|cc}
    \hline
    IIR  & \multicolumn{1}{c|}{IIH} & Cover / Stego  & Secret / Recovery\\
    \hline
    \hline
    8     & 8     & 32.76 / 0.865   & 33.48 / 0.940 \\%updated
    8     & 16    & 37.66 / 0.945   & 33.36 / 0.939 \\%updated
    16    & 8     & 32.56 / 0.861   & 33.72 / 0.941 \\%updated
    16    & 16    & 35.93 / 0.923   & 33.65 / 0.939 \\%updated
    Bicubic & 16  & 38.17 / 0.954   & 29.39 / 0.881 \\%updated
    \hline
    \end{tabular}%
  \label{tab:ablation_InvBlocks_layers}%
\end{table}%
%------------------------------------------------------------------------------

\textbf{Results on hiding $4$ images.} Table~\ref{tab:conceal_four_2} presents the results on hiding $4$ images.
We can observe that InvMIHNet achieves the best results among three methods on the quality of concealing and revealing on all three datasets. Specifically, InvMIHNet achieves 3.72 dB, 4.05 dB and 2.57 dB improvements on DIV2K, COCO and ImageNet datasets on concealing performance compared with DeepMIH~\cite{guan2022deepmih}, respectively. 
Meanwhile, InvMIHNet also surpasses ISN~\cite{Lu2021LargecapacityIS} by 5.71 dB, 2.45 dB and 4.97 dB on the revealing performance. 
% So more secret contents are precisely reconstructed. 
In terms of the computational complexity, InvMIHet takes about 5.44 seconds to conceal and reveal four images of  $1024 \times 1024$ with $5.71M$ learnable parameters. And the time consumption is only about the half of the two comparison methods. Meanwhile, the trainable parameters of DeepMIH~\cite{guan2022deepmih} and ISN~\cite{Lu2021LargecapacityIS} are $26.46M$ and $1.75M$, respectively.

\textbf{Results on hiding $N$ $(N>4)$ images.} 
Fig. \ref{fig:memory_consumption} shows the memory consumption of ISN, DeepMIH and InvMIHet on hiding different number of secret images. We can see that InvMIHNet has significantly lower memory consumption than DeepMIH~\cite{guan2022deepmih} and ISN~\cite{Lu2021LargecapacityIS}. We mainly compare with ISN on hiding $N>4$ images due to the memory limitation.
The performance of ISN~\cite{Lu2021LargecapacityIS} and InvMIHNet evaluated on the three datasets are shown in Fig.~\ref{fig:mult_image}, \textit{i.e.}, with $N=6,8,9,16$, which have never been achieved and reported by the other methods in the literature. 
We can consistently observe that InvMIHNet effectively improves the concealing and revealing performance compared to ISN~\cite{Lu2021LargecapacityIS}. 
Fig.~\ref{fig:8images} visualizes experimental results on hiding 8 images of ISN~\cite{Lu2021LargecapacityIS} and InvMIHNet on ImageNet. We can observe that the generated stego images and recovered secret images of ISN~\cite{Lu2021LargecapacityIS} suffer from serious color distortions, which results in difficulty in information separation and contents reconstruction.
Our proposed InvMIHNet can successfully hide more images with marginal increasing of number of parameters and hence significantly reduces the training resources compared to DeepMIH~\cite{guan2022deepmih}.

\subsection{Ablation study}
Invertible Blocks are the key building block of two Invertible Image modules. Table \ref{tab:ablation_InvBlocks_layers} shows the performances of InvMIHNet using different layers of IIR and IIH modules on hiding 4 images on DIV2K.
We can observe that increasing the number of InvBlocks in IIH module results in an improved quality of both concealing and revealing, and IIR module achieves comparable performance when the number of InvBlocks are 8 and 16.
We further replace IIR module with Bicubic interpolation. Without IIR module to preserve essential information into the downscaled part and supplement lost contents, the quality of recovery deteriorates quickly. 
This validates that IIR module plays a vital role in InvMIHNet and makes it possible to improve the hiding capacity and revealing quality simultaneously.
Considering the trade-off between performance and complexity, the default numbers of InvBlocks in IIR and IIH modules are set to 8 and 16, respectively.

\section{Conclusion}
\label{sec:conclusion}
In this paper, we introduce a novel and effective Invertible Mosaic Image Hiding Network (InvMIHNet) based on Invertible Neural Networks for very large capacity image stenography. 
The proposed InvMIHNet conceals the downscaled secret image into the the cover image leading to significant improvement on hiding capacity and recovery quality. 
Benefiting from the reversibility of the Invertible Image Rescaling module and Invertible Image Hiding module, the image concealing and revealing processes of InvMINHet are fully coupled and reversible. 
Extensive experimental results show that the proposed InvMIHNet outperforms state-of-the-art methods on the imperceptibility of stego images and recover accuracy of the secret images.
Moreover, the capability of InvMIHNet on very large capacity image steganography are further demonstrated with acceptable computational complexity and memory consumption.

\vfill\pagebreak

% \section{REFERENCES}
% \label{sec:refs}

% \small
\newpage
\bibliographystyle{IEEEbib}
\bibliography{refs}

\end{document}